\documentclass[12pt,a4paper]{article}
	
\usepackage{a4wide}
\usepackage{amsmath}
\usepackage{amssymb}
\usepackage{amsfonts}
\usepackage{epsfig}
\usepackage{exscale}
\usepackage{float}
\usepackage{bbm}
\usepackage[numbers,sort&compress]{natbib}

\newcommand{\R}{{\mathbb{R}}}

\newcommand{\p}{\partial}

\makeatletter
\@addtoreset{equation}{section}
\makeatother

\title{ Harmonic Oscillator in a 1D or 2D Cavity with General Perfectly Reflecting Walls}

\author{M.\ H.\ Al-Hashimi \\ \\
Albert Einstein Center for Fundamental Physics \\
Institute for Theoretical Physics, Bern University \\
Sidlerstrasse 5, CH-3012 Bern, Switzerland \\ \\}

\begin{document}

\maketitle

\vspace{-1cm}

\begin{abstract} \normalsize
We investigate the simple harmonic oscillator in a 1-d box, and the 2-d isotropic harmonic oscillator  problem in a circular cavity with perfectly reflecting
boundary conditions. The energy spectrum has been calculated as a function of the self-adjoint extension parameter. For sufficiently negative values of the
self-adjoint extension parameter, there are bound states localized at the wall of the box or the cavity that resonate with the standard bound states of the simple
harmonic oscillator or the isotropic oscillator. A free particle in a circular cavity has been studied for the sake of comparison. This work represents an application
of the recent generalization of the Heisenberg uncertainty relation related to the theory of self-adjoint extensions in a finite volume.
\end{abstract}

\newpage

\section{Introduction}
Studying quantum systems confined to a finite volume is not a new problem. There is a collection of articles studying this subject \cite{Mic37,Som38,Gro46,Wig54,Fow84,Fro87,Shaqqur1,Shaqqur2,Sch00,Vos,Shuai,Pup98,Pup02}. These articles cover especially the hydrogen atom in a finite volume. In addition, there are articles that cover the isotropic harmonic oscillator in a finite volume \cite{Sab09,LeSar,Art,Marin,Aqu,Sen1,Sen2,Sen3}. Studying the isotopic harmonic oscillator energy spectrum in confinement falls in line with contemporary
applications in the areas of mesoscopic scale semiconductor structures like quantum dots containing one to a few electrons. The isotropic harmonic oscillator has been used as a model to study the molecular vibration spectrum in solids, and the magnetic properties of an electron gas confined in a semiconductor structure \cite{Har05,Sen2}. In all the previous studies of this problem, it was assumed that the wave function vanishes at the boundaries. This condition is known as the Dirichlet boundary condition. \\ In a previous paper \cite{AlH12}, we studied the problem of a particle in a box not by assuming that the wave function necessarily vanishes at the boundaries, but by assuming that there is no probability leaking outside the box. This means that the probability density at the boundaries is not necessarily zero. This is consistent with the classical intuition of a ball bouncing off a perfectly reflecting wall. The condition at the boundaries is dictated by demanding that the Hamiltonian is self-adjoint. This boundary condition is known as a Robin boundary condition.  The results of \cite{AlH12} were applied to the free particle and the hydrogen atom in 3-d  in our previous article  \cite{PaperNo.5}. In this work, we study the free particle in a circular cavity in 2-d, the simple harmonic oscillator in  1-d, and the isotropic harmonic oscillator in 2-d by using the general Robin boundary condition. \\The totally reflecting boundaries could have an infinite number of features, because  there is an infinite number of potentials at the wall that can make the wall totally reflective. The feature of a boundary influences the energy spectrum, and this influence can be described  elegantly by a self-adjoint extension parameter \cite{AlH12}. One of the motivations for this article is to understand the relation between the energy spectrum of the simple harmonic oscillator in a 1-d box, and a 2-d isotropic oscillator in a circular cavity  and the value of the self-adjoint extension parameter. The free particle in 2-d is studied for the sake of comparison with the 2-d isotropic harmonic oscillator. \par Now consider a self-adjoint extension parameter $\gamma(\vec x)$ that specifies the physical properties of the reflecting wall \cite{AlH12}. The corresponding general Robin boundary condition takes the form
\begin{equation}
\label{bcdot}
\gamma(\vec x) \Psi(\vec x) + \vec n(\vec x) \cdot \vec \nabla \Psi(\vec x) = 0,
\quad \vec x \in \p \Omega,
\end{equation}
where $\p \Omega$ is the boundary of a spatial region $\Omega$, and $\vec n(\vec x)$ is the unit-vector normal to the surface. In the  textbook case one sets $\gamma(\vec x) = \infty$, which implies that the wave function $\Psi(\vec x)$ vanishes at the boundary. However, this is not necessarily the case because eq.(\ref{bcdot}) always guarantees that
\begin{equation}\label{neq}
\vec n(\vec x) \cdot \vec j(\vec x) = 0, \quad \vec x \in \p \Omega,
\end{equation}
where the current density $j(\vec x)$ is defined by the following equation
\begin{equation}
\vec j(\vec x,t) = \frac{1}{2 M i}
\left[\Psi(\vec x,t)^* \vec \nabla \Psi(\vec x,t) -
\vec \nabla \Psi(\vec x,t)^* \Psi(\vec x,t)\right].
\end{equation}
Eq.(\ref{neq}) ensures that the component of $\vec j(\vec x)$  normal to the surface vanishes. Hence, together with the continuity equation
\begin{equation}
\p_t \rho(\vec x,t) + \vec \nabla \cdot \vec j(\vec x,t) = 0, \quad
\rho(\vec x,t) = |\Psi(\vec x,t)|^2,
\end{equation}
the boundary condition eq.(\ref{bcdot}) ensures probability conservation.
This is the key feature that guarantees the self-adjointness (rather than just
the Hermiticity) of the Hamiltonian
\begin{equation}
H = \frac{{\vec p \,}^2}{2 M} + V(\vec x) = - \frac{1}{2 M} \Delta + V(\vec x).
\end{equation}
To understand the delicate issue of Hermiticity versus self-adjointness, let us first consider
\begin{eqnarray}
\langle\chi|H|\Psi\rangle&=&
\int_\Omega d^dx \ \chi(\vec x)^*
\left[- \frac{1}{2 M} \Delta + V(\vec x)\right] \Psi(\vec x) \nonumber \\
&=&\int_\Omega d^dx \
\left[\frac{1}{2 M} \vec \nabla \chi(\vec x)^* \cdot \vec \nabla \Psi(\vec x) +
\chi(\vec x)^* V(\vec x) \Psi(\vec x) \right] \nonumber \\
&-&\frac{1}{2 M}
\int_{\p \Omega} d\vec n \cdot \chi(\vec x)^* \vec \nabla \Psi(\vec x)
\nonumber \\
&=&\int_\Omega d^dx \ \left\{\left[- \frac{1}{2 M} \Delta + V(\vec x)\right]
\chi(\vec x)^* \right\} \Psi(\vec x) \nonumber \\
&+&\frac{1}{2 M} \int_{\p \Omega} d\vec n \cdot
\left[\vec \nabla \chi(\vec x)^* \Psi(\vec x) -
\chi(\vec x)^* \vec \nabla \Psi(\vec x)\right] \nonumber \\
&=&\langle\Psi|H|\chi\rangle^* + \frac{1}{2 M} \int_{\p \Omega} d\vec n \cdot
\left[\vec \nabla \chi(\vec x)^* \Psi(\vec x) -
\chi(\vec x)^* \vec \nabla \Psi(\vec x)\right].
\end{eqnarray}
The Hamiltonian is Hermitian if
\begin{equation}
\label{symmetricHdot}
\int_{\p \Omega} d\vec n \cdot
\left[\vec \nabla \chi(\vec x)^* \Psi(\vec x) -
\chi(\vec x)^* \vec \nabla \Psi(\vec x)\right] = 0.
\end{equation}
Using the boundary condition eq.(\ref{bcdot}), the integral in
eq.(\ref{symmetricHdot}) reduces to
\begin{equation}
\int_{\p \Omega} d^{d-1}x \left[\vec n(\vec x) \cdot \vec \nabla \chi(\vec x)^* +
\gamma(\vec x) \chi(\vec x)^*\right] \Psi(\vec x) = 0.
\end{equation}
Since $\Psi(\vec x)$ itself can take arbitrary values at the boundary, the
Hermiticity of $H$ requires that
\begin{equation}
\label{bcdual}
\vec n(\vec x) \cdot \vec \nabla \chi(\vec x) + \gamma(\vec x)^* \chi(\vec x) =
0.
\end{equation}
For $\gamma(\vec x) \in \R$, this is again the boundary condition of
eq.(\ref{bcdot}), which ensures that $D(H^\dagger) = D(H)$, such that $H$ is
indeed self-adjoint.

\section{The Energy Spectrum of the Simple Harmonic Oscillator in 1-D Confinement}
\begin{figure}[tbh]
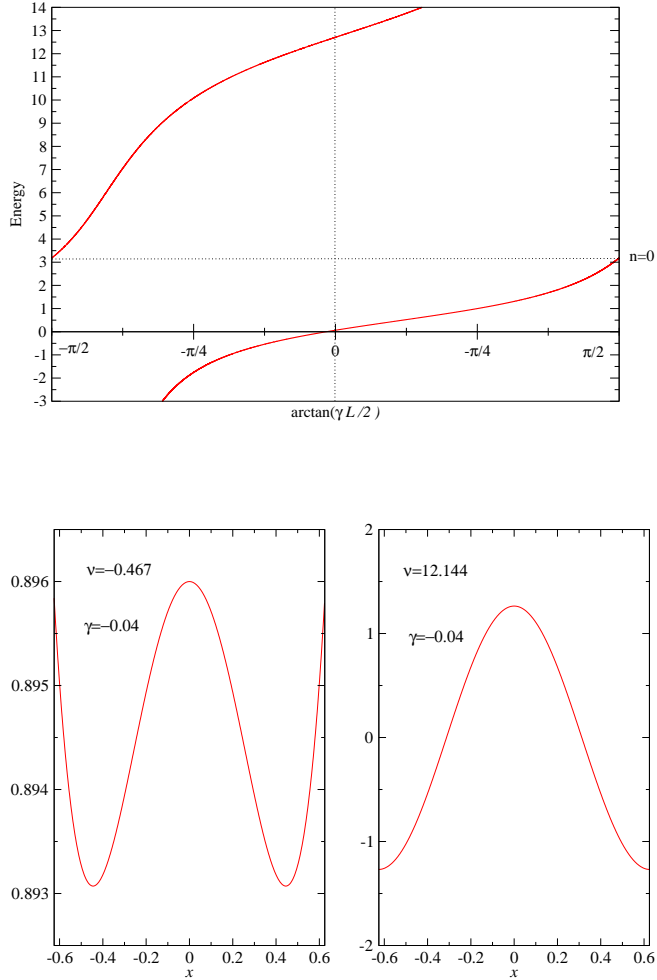

\vskip1 cm
\begin{center}
\epsfig{file=iso5.eps,width=8.5cm} \vskip1.3cm
\epsfig{file=iso6.eps,width=8.5cm}
\end{center}
\caption{\it Top: Avoided level crossing between the lowest two states for even wave functions of the simple harmonic operator centered in a 1-d box
of width $L =1.25 \alpha^{-1}$ indicating a cavity resonance. The energy
is given in units of $\omega$. Bottom: Wave functions of the two states for
$\gamma = -0.04 \alpha$, which are localized both near the center and at the walls
of the box.}
\label{isofig3}
\end{figure}

In 1-d,  eq.(\ref{bcdot}) can be written as
\begin{equation}\label{boundarySHOgen}
\gamma(x) \Psi(x) + \p_x \Psi(x) = 0, \quad x = \pm L/2,
\end{equation}
where $L$ is the size of the finite interval in which the particle is confined. The two real-valued parameters $\gamma(L/2)$ and $\gamma(-L/2)$
represent  one parameter  families of self-adjoint extensions of the Hamiltonian at each of the
two ends of the interval $\Omega$. In order not to break parity via the boundary
conditions, the following restriction must be taken into consideration \cite{AlH12}
\begin{equation}
\gamma(L/2) = - \gamma(-L/2) = \gamma \in \R,
\end{equation}
such that
\begin{equation}
\label{bc}
\gamma \Psi(L/2) + \p_x \Psi(L/2) = 0, \quad
- \gamma \Psi(-L/2) + \p_x \Psi(-L/2) = 0.
\end{equation}
Consider a simple harmonic oscillator with the center of force located at the center of a 1-d box. The Schr\"{o}dinger equation for a particle in an  energy eigenstate state is \cite{shiff}
\begin{equation}\label{SchSHO}
     E \Psi(x)=-\frac{1}{2 M}\frac{\partial^2 \Psi(x)}{\partial x^2}+\frac{1}{2}k x^2 \Psi(x),
\end{equation}
where $k$ is Hooke's constant, and  $M$ is the mass of the particle. The general solution for the above equation is
\begin{equation}\label{SHOsolution1}
    \Psi_\nu(x)=A D_{\nu}(\sqrt{2}\alpha x)+ B D_\nu( -\sqrt{2}\alpha x),
\end{equation}
where $D_{\nu}(\sqrt{2}\alpha x)$ is the parabolic cylindrical function \cite{wolfram}, $A$ and $B$ are constants, and $\alpha=\sqrt[4]{M k}$. The energy spectrum is given by the following equation.
\begin{equation}\label{EnSHO}
    E=\omega \left(\frac{1}{2}+\nu\right), \hspace{3mm} \omega=\sqrt{\frac{k}{M}},
\end{equation}
Bear in mind that the value of $\nu$ needn't be an integer. We have to remember that in the infinite volume, $\nu$ was put to an integer to ensure that the wave function vanishes at infinities. In fact, for integer $\nu$, the parabolic cylindrical function can be written as \cite{wolfram}
\begin{equation}
    D_{n}(\sqrt{2}x)=2^{-\frac{n}{2}}\exp(-\frac{x^2}{2})H_n(x),
\end{equation}
 where $ H_n(x)$ is the Hermite function that appears in the expression of the simple harmonic oscillator wave function in the infinite volume.
 \\ For the wave function in eq.(\ref{SHOsolution1}) to be either an odd or even, $A$ and $B$  must be restricted. For an even wave function $B=A$, and for an odd wave function $B=-A$.\\
For even states, the wave function is written as
\begin{equation}\label{evenSHO}
     \Psi_\nu(x)=A (D_{\nu}(\sqrt{2}\alpha x)+  D_\nu( -\sqrt{2}\alpha x))
\end{equation}
By substituting eq.(\ref{evenSHO}) in the boundary condition of eq.(\ref{boundarySHOgen}), we get the transcendental equation that determines the spectrum for this case, which is
\begin{equation}\label{gamaIeven}
  \left(\frac{L\alpha^2}{2}+\gamma\right)\left(D_{\nu}(- \frac{L\alpha}{\sqrt{2}})+D_{\nu}( \frac{L\alpha}{\sqrt{2}})\right)+\sqrt{2}\alpha\left(D_{\nu+1}(- \frac{L\alpha}{\sqrt{2}})-D_{\nu+1}( \frac{L\alpha}{\sqrt{2}})\right)=0.
\end{equation}
 For odd states, the wave function is written as
\begin{equation}\label{oddSHO}
     \Psi_\nu(x)=A (D_{\nu}(\sqrt{2}\alpha x)-  D_\nu( -\sqrt{2}\alpha x))
\end{equation}
By substituting eq.(\ref{oddSHO}) in the boundary condition of eq.(\ref{boundarySHOgen}), we get
\begin{equation}\label{gamaIodd}
  \left(\frac{L\alpha^2}{2}+\gamma\right)\left(-D_{\nu}(- \frac{L\alpha}{\sqrt{2}})+D_{\nu}( \frac{L\alpha}{\sqrt{2}})\right)+\sqrt{2}\alpha\left(D_{\nu+1}(- \frac{L\alpha}{\sqrt{2}})+D_{\nu+1}( \frac{L\alpha}{\sqrt{2}})\right)=0.
\end{equation}
\begin{figure}[H]
\begin{center}
\epsfig{file=iso1.eps,width=10cm} \vskip0.5cm
\epsfig{file=iso2.eps,width=10cm}
\end{center}
\caption{\it Top: Spectrum of the simple harmonic oscillator with even wave functions in a 1-d box
of width $L =5 \alpha^{-1}$ with general Robin boundary conditions as a
function of the self-adjoint extension parameter $\gamma$, rescaled to
$\arctan(\gamma L/2)$. The energy is given in units of $\omega$.  The dotted
lines represent the spectrum for $\gamma = \infty$. Bottom: Wave functions of
the four lowest even states with $n = 0, 1, 2$, and $3 $ for
$\gamma = \infty,0,-2/L$, and $-\infty$.}
\label{isofig1}
\end{figure}
\begin{figure}[H]
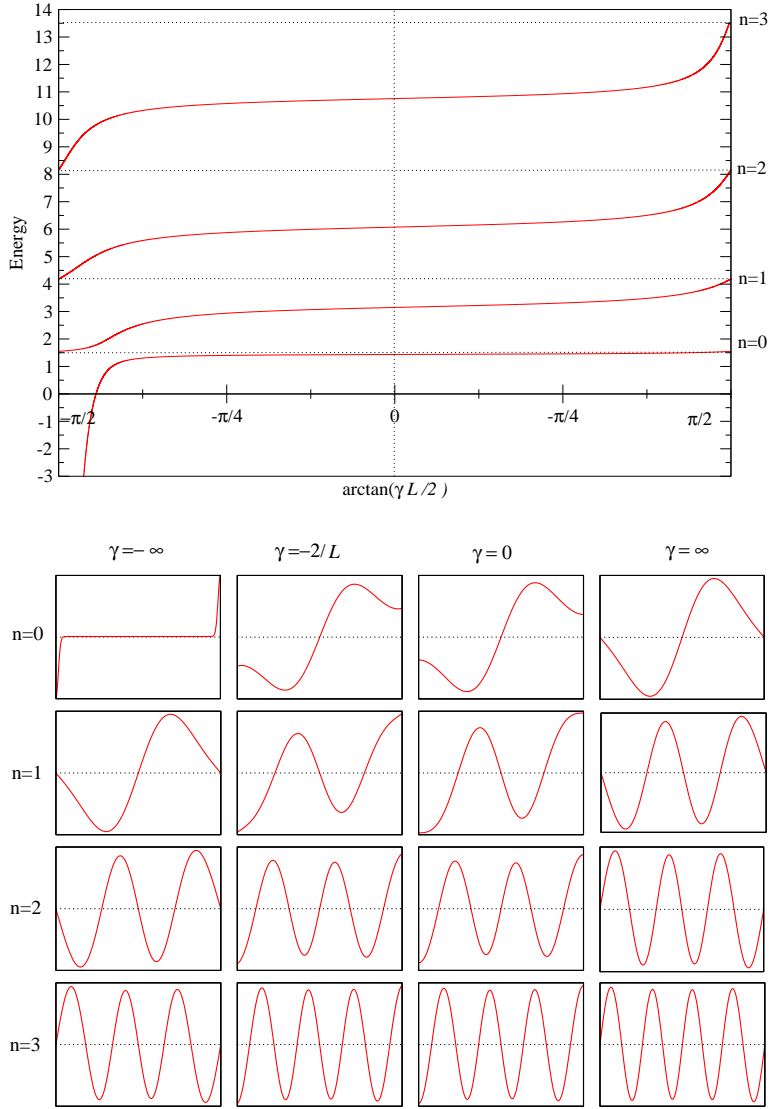

\begin{center}
\epsfig{file=iso3.eps,width=10cm} \vskip0.5cm
\epsfig{file=iso4.eps,width=10cm}
\end{center}
\caption{\it Top: Spectrum of the simple harmonic oscillator with odd wave functions in a 1-d box
of width $L =5 \alpha^{-1}$  with general Robin boundary conditions as a
function of the self-adjoint extension parameter $\gamma$, rescaled to
$\arctan(\gamma L/2)$. The energy is given in units of $ \omega$.  The dotted
lines represent the spectrum for $\gamma = \infty$. Bottom: Wave functions of
the four lowest odd states with $n = 0, 1, 2$, and $3$ for
$\gamma = \infty,0,-2/L$, and $-\infty$.}
\label{isofig2}
\end{figure}
For even states, the energy spectrum as a function of the self-adjoint extension parameter
$\gamma$ and the corresponding wave functions of the states with
$n = 0, 1, 2,$ and $3$ are illustrated in figure \ref{isofig1}. For odd states,  analogous results are shown in figure \ref{isofig2}.
In both of these figures at the top, we notice avoided level crossings, which correspond to simple harmonic oscillator
states that resonate with states localized at the  walls due to the introduction of the Robin boundary condition. Figure \ref{isofig3}
magnifies the resonance effect by reducing the width of the box to $L=1.25 \alpha^{-1}$.
The avoided level crossing is more obvious between the ground state and the first  excited state. The corresponding wave functions are illustrated at the bottom of figure \ref{isofig3}. It is obvious from the bottom of figure \ref{isofig3} that a cavity resonance occurs for a negative value of $\nu$, which corresponds to a certain value of
$\gamma$. By using numerical methods, and calculating the values of $\gamma$ for which the wave function shows a similar shape as the one at the bottom of  figure \ref{isofig3}, for different box sizes we obtain table 1 for even states.
\begin{table}
\begin{center}
\begin{tabular}{|c|c|c|c|}
\hline
 $L$& $\gamma$ & $\nu$ & $E$\\\noalign{\hrule height 2pt}
 8 & -3.7288 & -0.0014 & 0.49858 \\\hline
7 & -3.1715 & -0.00795 & 0.49205 \\\hline
6 & -2.5657 & -0.0334& 0.4666\\\hline
5 & -1.8735 &-0.0962&0.4038\\\hline
4 & -1.1366& -0.1985&0.3015\\\hline
3 & -0.5349 & -0.3192&0.1808\\\hline
2 & -0.1641 & -0.4168&0.0832\\\hline
1 & -0.0206 & -0.4790&0.021\\\hline

\end{tabular}
\end{center}
\caption{\it From the left: $L$ is the width of the box in units of $\alpha^{-1}$, $\gamma$ is in units of $\alpha$, $\nu$ is the main quantum number, and $E$ to the right
is the resonance energy for the ground state in units of $\omega$.}
\end{table}
\section{Free Particle in a Circular Cavity with General Reflecting Boundaries}
To put the problem of the oscillator in good prospective, we consider first the problem of a free
particle in a circular cavity with general reflecting boundary conditions
specified by the self-adjoint extension parameter $\gamma \in \R$.
\subsection{Energy spectrum}
\begin{figure}[tbh]
\begin{center}
\epsfig{file=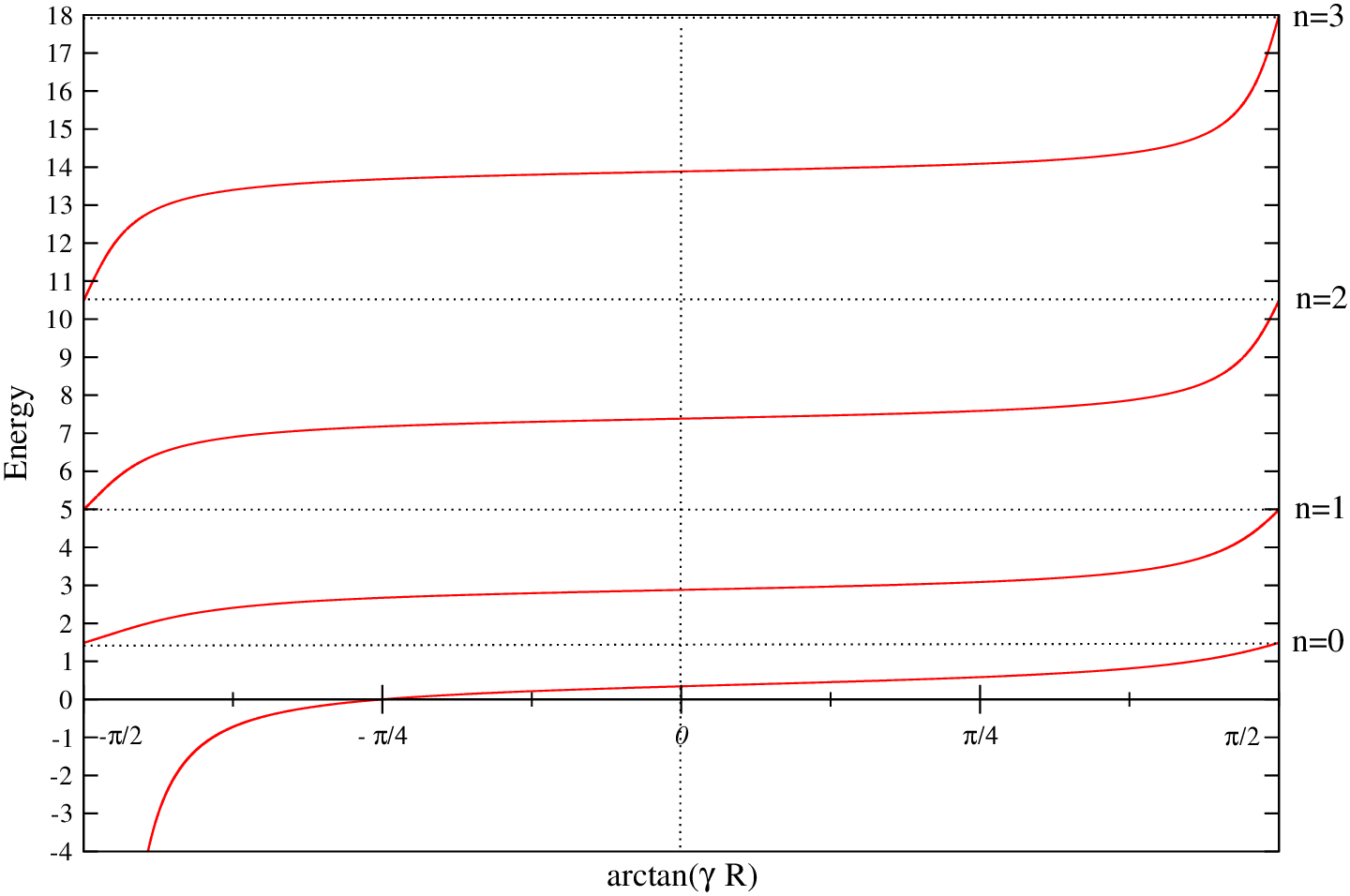,width=10cm} \vskip0.5cm
\epsfig{file=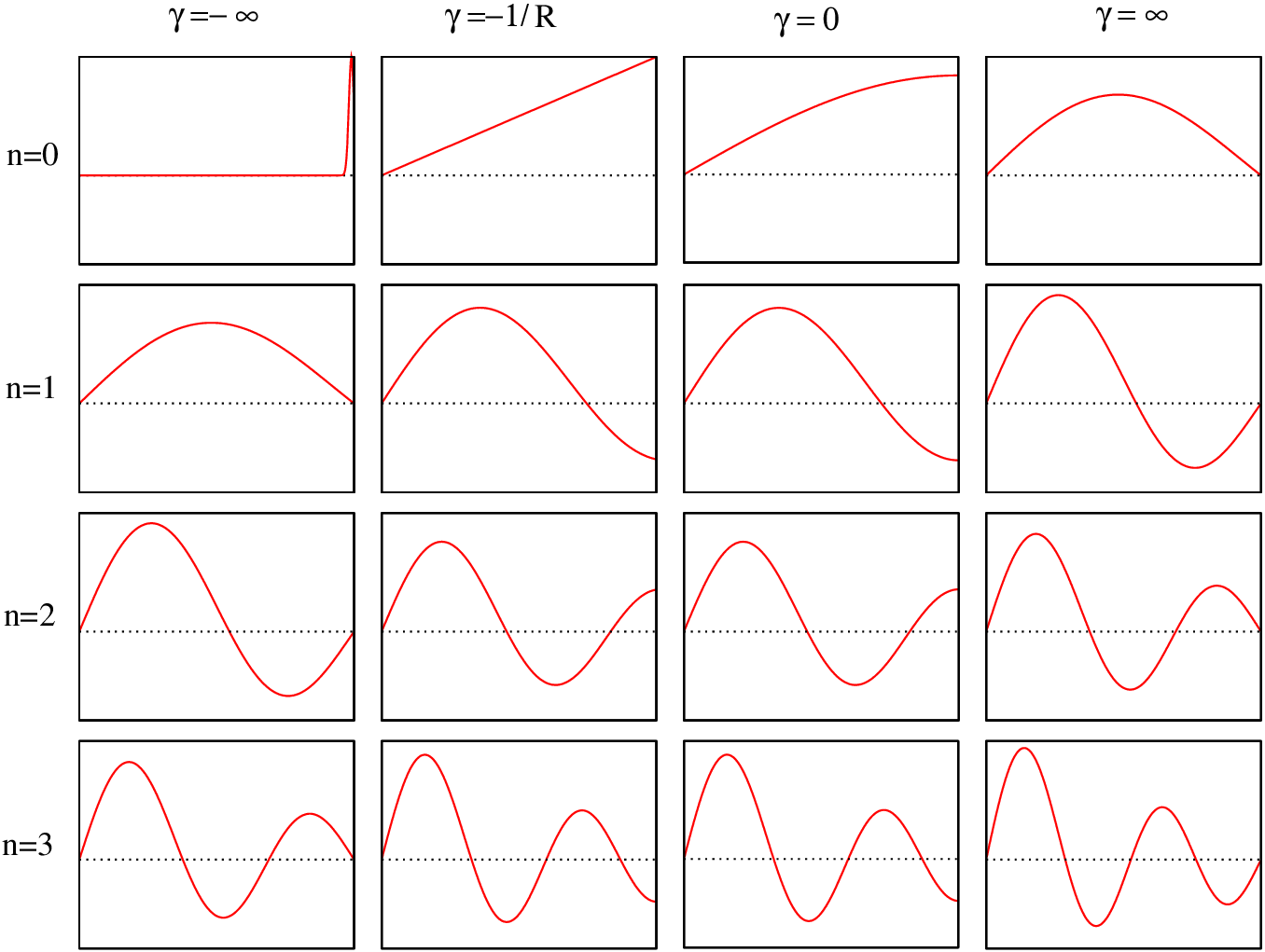,width=10cm}
\end{center}
\caption{\it Top: Spectrum of $|m| = 1$ states for a free particle in a circular
cavity with general boundary conditions as a function of the self-adjoint
extension parameter $\gamma$, rescaled to $\arctan(\gamma R)$. The energy
is measured in units of $\pi^2/2 M R^2$. The dotted lines represent the spectrum
for $\gamma = \infty$. Bottom: Wave functions of the four lowest $|m| = 1$ states
with $n = 0, 1, 2$ and $3 $ for $\gamma = \infty,0,-1/R$, and $-\infty$.}
\label{free1}
\end{figure}

The Hamiltonian of a free particle of mass $M$ is,
\begin{equation}
H = - \frac{1}{2 M} \Delta = - \frac{1}{2 M} \left(\p_r^2 + \frac{1}{r} \p_r
- \frac{ L^2}{r^2}\right),
\end{equation}
with angular momentum $L_z$ in a circular cavity of radius $R$. As usual,
the wave function can be factorized as
\begin{equation}
\Psi(\vec x) = \psi_{\nu m}(r) \exp(i m \varphi),  \hspace{5mm} m=0,\pm 1,\pm 2,...,
\end{equation}
where the angular dependence is described by the function $\exp(i m \varphi)$, and
 the radial wave function $ \psi_{\nu m}(r)$ satisfies the following relation
\begin{equation}
- \frac{1}{2 M} \left(\p_r^2 + \frac{1}{r} \p_r
- \frac{m^2}{r^2}\right) \psi_{\nu m}(r) = E \psi_{k m}(r), \quad
E = \frac{k^2}{2 M}.
\end{equation}
For positive energy, there are two linearly independent solutions to the above differential equation. Only one is finite at the origin. It takes the form
\begin{equation}
\psi_{k l} = A J_{|m|}(k r),
\end{equation}
where $J$ is the $J$-Bessel function. For a spherical cavity, the most general perfectly reflecting boundary condition of eq.(\ref{bcdot}) takes the form
\begin{equation}
\gamma \psi_{k m}(R) + \p_r  \psi_{k m}(R) = 0,
\end{equation}
The energy spectrum is thus determined from
\begin{equation}
\gamma J_{|m|}(k R) + \p_r  J_{|m|}(k R) =
\left( \gamma -\frac{|m|}{R} \right) J_{|m|}(k R) + k J_{|m|-1}(k R) = 0,
\end{equation}
which is a transcendental equation for $k \in \R$.
Let us consider  $\gamma \rightarrow \infty$. The boundary condition then reduces to the textbook case $\psi(R) = 0$. The
corresponding energies are then given by the roots of the $J$-Bessel function, i.e.
\begin{equation}
 J_{|m|}(k R)=0.
\end{equation}
Unlike in the 3-d case, in 2-d, there are no simple expressions for the energy of a particle for special values of $\gamma$.
An interesting case is $\gamma =-|m|/R$. Then, the energy spectrum can be evaluated from the equation
\begin{equation}
k J_{|m|+ 1}(k R)=0,
\end{equation}
The above equation is satisfied for zero energy states ($k=0$). The above equation also tells us that for $k\neq 0$, the energy spectrum for $\gamma = -|m|/R$ is the same as the energy spectrum for the textbook case $(\gamma\rightarrow\infty)$ with angular momentum quantum number $|m|+1$. For example, for the case $m=0$, the energy spectrum for $\gamma =0$ is the same as the energy spectrum for
$\gamma\rightarrow\infty$ with $|m|=1$.\\
Interestingly, for $\gamma < -|m|/R$, there are even negative energy states, although the particle only has kinetic
energy. This is a consequence of the general Robin boundary conditions. While they are perfectly reflecting for positive energy states, they may still bind
negative energy states to the wall. The negative energy states simply follow by analytic continuation of $k$ to $i k$. For $\gamma \rightarrow - \infty$
there is a bound state for each angular momentum $m$, with the energy
\begin{equation}
E_{0m} \rightarrow - \frac{\gamma^2}{2 M}
\end{equation}
\begin{figure}[tbh]
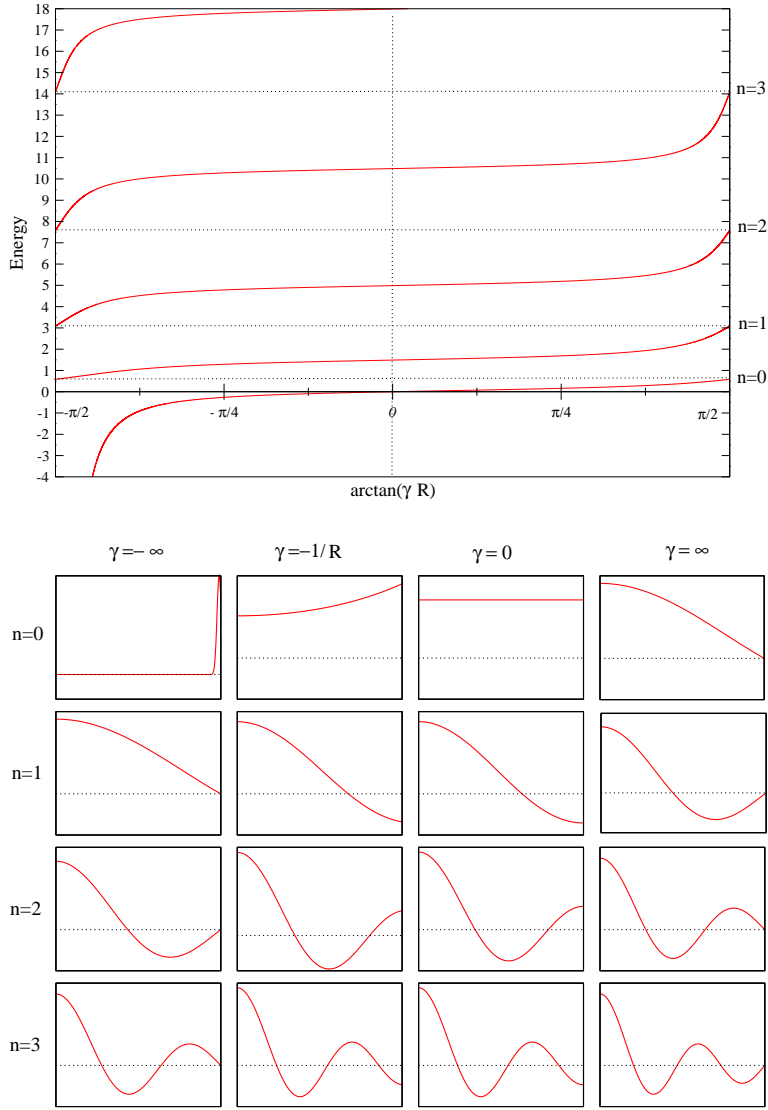

\begin{center}
\epsfig{file=iso7.eps,width=10cm} \vskip0.5cm
\epsfig{file=iso8.eps,width=10cm}
\end{center}
\caption{\it Top: Spectrum of $m = 0$ states for a free particle in a circular
cavity with general boundary conditions as a function of the self-adjoint
extension parameter $\gamma$, rescaled to $\arctan(\gamma R)$. The energy
is measured in units of $\pi^2/2 M R^2$. The dotted lines represent the spectrum
for $\gamma = \infty$. Bottom: Wave functions of the four lowest $m = 0$ states
with $n = 0, 1, 2$, and $3$ for $\gamma = \infty,0,-1/R$, and $-\infty$.}
\end{figure}
Analogous results are shown in figure \ref{free1} for $|m| =1$.
\subsection{Generalized Uncertainty Relation}
It is counter intuitive to have a free particle with negative energy when it is confined to a circular cavity with general Robin boundary conditions. This apparently
contradicts the Heisenberg uncertainty relation, because the necessarily finite
uncertainty of the position of a confined particle seems to imply a positive
kinetic energy. However, this is not necessarily the case because the standard
Heisenberg uncertainty relation was derived for an infinite volume and thus does not apply
in a finite cavity. We have derived a generalized uncertainty relation
valid in an arbitrarily shaped finite region $\Omega$ with the unit-vector
$\vec n$ perpendicular to the boundary $\p \Omega$ \cite{AlH12}
\begin{equation}
\label{uncertainty}
2 M E_n = \langle {\vec p \,}^2 \rangle \geq \left(\frac{2 +
\langle \vec n \rangle \cdot \langle \vec x \rangle -
\langle \vec n \cdot \vec x \rangle}{2 \Delta x}\right)^2 +
\langle \gamma \rangle +
\frac{\langle \vec n \rangle^2}{4},
\end{equation}
where we have defined
\begin{eqnarray}
&&\langle \vec n \cdot \vec x \rangle =
\int_{\p \Omega} d\vec n \cdot \vec x \rho(\vec x), \nonumber \\
&&\langle \gamma \rangle =
\int_{\p \Omega} ds  \gamma(\vec x) \rho(\vec x), \nonumber \\
&&\langle \vec n \rangle =
\int_{\p \Omega} d\vec n \ \rho(\vec x), \quad
\rho(\vec x) = |\Psi(\vec x)|^2.
\end{eqnarray}
Here $ds$ is a length element of the circular cavity. In the infinite volume limit, for localized states (with momentum expectation
value $\langle \vec p \rangle = 0$), the probability density
vanishes at infinity, and we obtain $\langle \vec n \cdot \vec x \rangle = 0$,
$\langle \gamma \rangle = 0$, and $\langle \vec n \rangle = 0$, such that we
recover the usual Heisenberg uncertainty relation in two dimensions
\begin{equation}
\langle {\vec p \,}^2 \rangle \geq \frac{1}{ (\Delta x)^2} \
\Rightarrow \ \Delta x \Delta p \geq 1.
\end{equation}
For a free particle confined to a circular cavity in an energy eigenstate, we obtain
\begin{equation}
\langle \vec n \cdot \vec x \rangle = R |\psi_{k m}|^2, \quad
\langle \gamma \rangle = \gamma R |\psi_{k m}|^2, \quad
\langle \vec n \rangle = 0.
\end{equation}
We are particularly interested in the zero-energy states, i.e.\
$k \rightarrow 0$, which arise for $\gamma = -|m|/R$. In that case, we obtain
\begin{equation}\label{HR}
\langle \vec n \cdot \vec x \rangle = 2|m|+ 2, \quad
\langle \gamma \rangle =  (2|m| + 2) \frac{\gamma}{R} = -\frac{|m|(2 |m| +2)}{R^2},
\quad \Delta x = R\sqrt{\frac{|m| + 1}{|m| + 2}}.
\end{equation}
Substituting eqs.(\ref{HR}) in the generalized uncertainty relation eq.(\ref{uncertainty}) gives
\begin{equation}
0 \geq \frac{1}{R^2}\left(\frac{|m|^2( |m| + 2)}{|m| + 1} - 2|m|(|m| + 1)\right).
\end{equation}
The above inequality is satisfied for all values of $|m|>0$, because
\begin{equation}
0 >  -  |m|^2-2|m|-2.
\end{equation}
For $m= 0$ the inequality is saturated. This means that the corresponding wave
function, which is constant for $\gamma = 0$, represents a minimal uncertainty wave packet in the finite volume. A similar result has been found for a free particle in 3-d \cite{PaperNo.5}
\section{An Isotropic Harmonic Oscillator in a Circular Cavity with General Reflecting Boundaries}
In this section, we consider the problem of an isotropic harmonic oscillator in a circular cavity  with general reflecting boundary conditions, where the center of the potential is at the center of the cavity. Again here we specify the self-adjoint extension parameter  as $\gamma \in \R$.
\subsection{Energy spectrum}
For the case of the isotropic harmonic oscillator in 2-d, the Hamiltonian takes the following form,
\begin{equation}\label{ISOSch}
H = - \frac{1}{2 M} \Delta +\frac{1}{2} k r^2=
- \frac{1}{2 M} \left(\p_r^2 + \frac{1}{r} \p_r
- \frac{ L^2}{r^2}\right) +\frac{1}{2} k r^2.
\end{equation}
The wave function can be factorized into a radial part and an angular part
\begin{equation}
\Psi(\vec x) = \psi_{\nu m}(r) \exp(i m \varphi),
\end{equation}
and the radial part of the Schr\"{o}dinger equation for the Hamiltonian of eq.(\ref{ISOSch}) is
\begin{equation}\label{ISORadial}
\left[- \frac{1}{2 M} \left(\p_r^2 + \frac{1}{r} \p_r
- \frac{m^2}{r^2}\right) +\frac{1}{2}k r^2\right] \psi_{\nu m}(r) =
E \psi_{\nu m}(r).
\end{equation}
In this case, the energy is
\begin{equation}
E =  \omega(\nu+1).
\end{equation}
Unlike in the infinite volume case, here $\nu$ is not necessarily a positive integer. It can take any negative or positive real value. It is useful to remember that in an infinite volume the energy can be written as $ E= \omega(2 n_r+|m|+1)$,  where $n_r$ is the radial quantum number, which is a positive integer. It is obvious that in an infinite volume, $m$ and $n_r$ can take different values corresponding to the same energy level. That is why the the energy levels are degenerate. This degeneracy is removed in general in a finite volume.\\By solving eq.(\ref{ISORadial}) we get two linearly independent  solutions. One of them is discarded because it is infinite at the origin. Accordingly, the solution of eq.(\ref{ISORadial}) is
\begin{equation}\label{ISOWF}
  \psi_{\nu m}(r) =A \exp(-\frac{\alpha^2r^2}{2})(\alpha r)^{|m|} \hspace{0.5mm}{}_1F_1(\tfrac{1}{2}(|m|-\nu), |m|+1,\alpha^2 r^2),
\end{equation}
where ${}_1F_{1}$ is the confluent hypergeometric function \cite{Arfken}. \\As
before, for a circular cavity with the most general perfectly reflecting
boundary condition we have
 \begin{equation}\label{boundaryISO}
\gamma \psi_{\nu m}(R) + \p_r  \psi_{\nu m}(R) = 0,
\end{equation}\label{RobinBC}
By substituting eq.(\ref{ISOWF}) in the boundary condition of eq.(\ref{boundaryISO}), we get the transcendental equation that determines the spectrum in this case. The equation is
 \begin{eqnarray}\label{IsoSpectrum}
  (1+|m|)  \left(R \alpha^2-\frac{|m|}{R}-\gamma\right) \hspace{0.5mm}{}_1\widetilde{F}_{1}(\tfrac{1}{2}(|m|-\nu),|m|+1,R^2 \alpha^2)&+&\nonumber\\ R \alpha^2(\nu-|m|)\hspace{0.5mm}{}_1\widetilde{F}_1(\tfrac{1}{2}(|m|+2-\nu),|m|+2,R^2 \alpha^2)&=&0,
 \end{eqnarray}
where ${}_1\widetilde{F}_{1}$ is the regularized hypergeometric function \cite{wolfram}.

\begin{figure}[tbh]
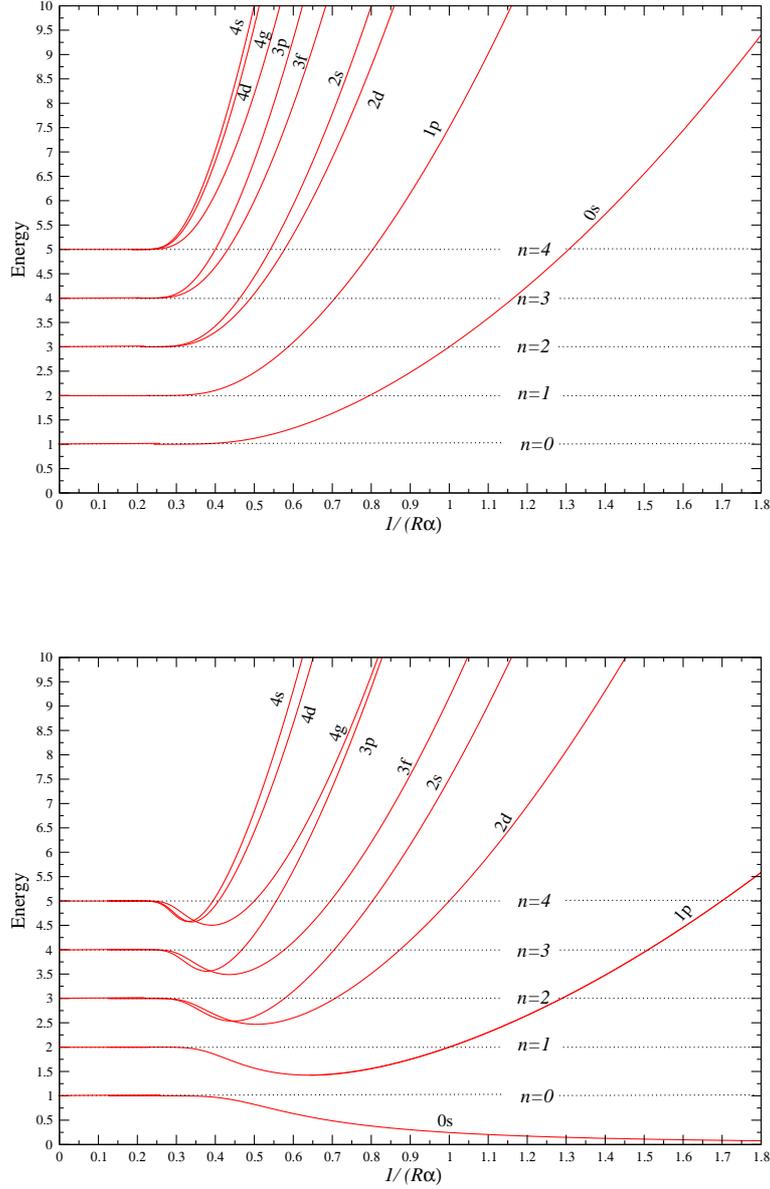

\begin{center}
\epsfig{file=iso11.eps,width=10cm} \vskip1.5cm
\epsfig{file=iso12.eps,width=10cm}
\end{center}
\caption{\it Top: Spectrum of the isotropic harmonic oscillator centered in a  circular cavity with
Dirichlet boundary condition (i.e.\ $\gamma = \infty$) as a function of $1/(R\alpha)$. The energy is in units of $\omega$. The
dotted lines represent the spectrum of the infinite system. Bottom: Spectrum of the isotropic harmonic oscillator centered in a  circular cavity with
Neumann boundary condition (i.e.\ $\gamma = 0$). Here, s, p, d, f, and g stand for states with $|m|=0,1,2,3,$ and $4$ respectively.}
\label{isofig4}
\end{figure}

\begin{figure}[tbh]
\begin{center}
\epsfig{file=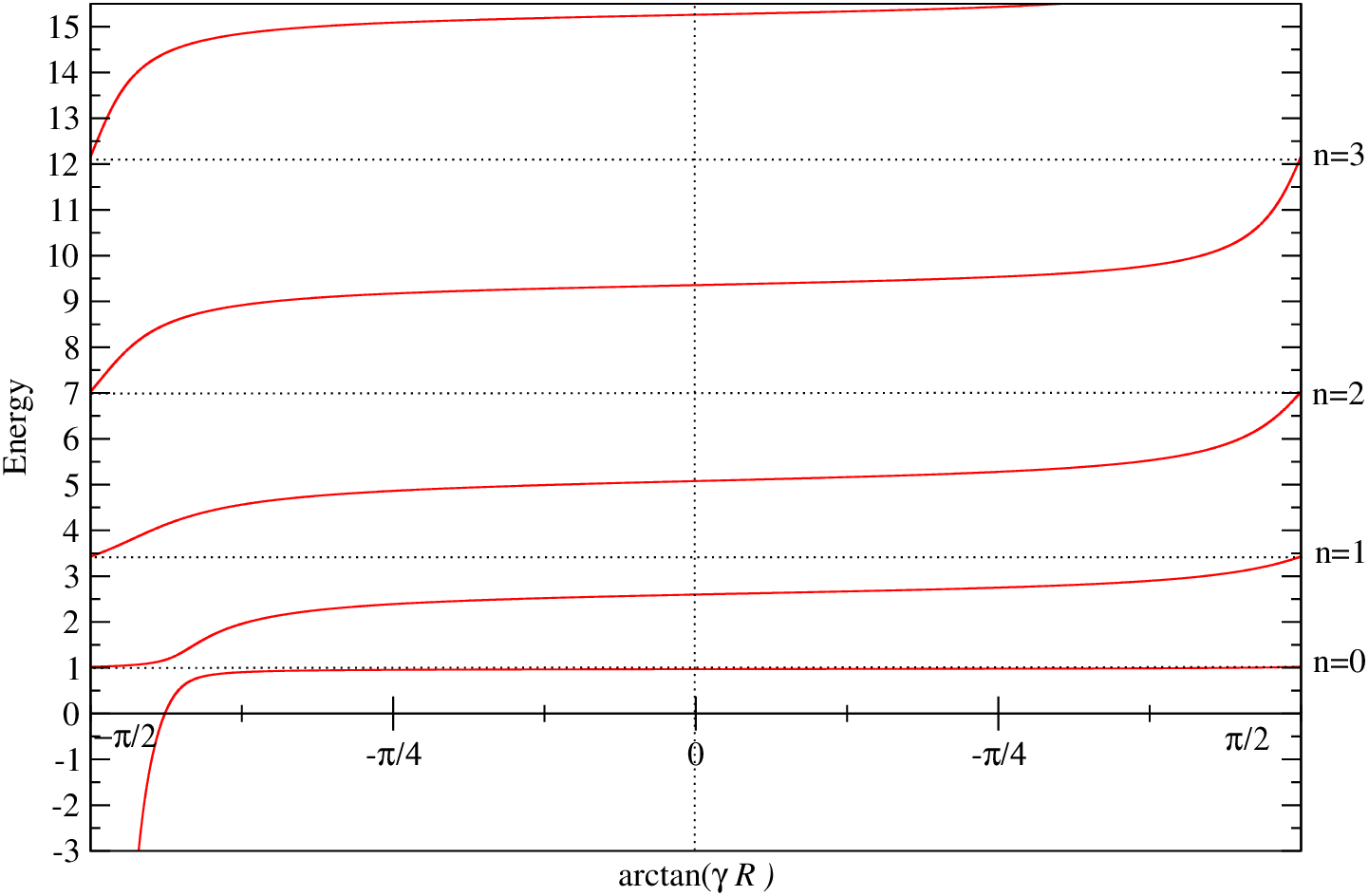,width=10cm} \vskip0.5cm
\epsfig{file=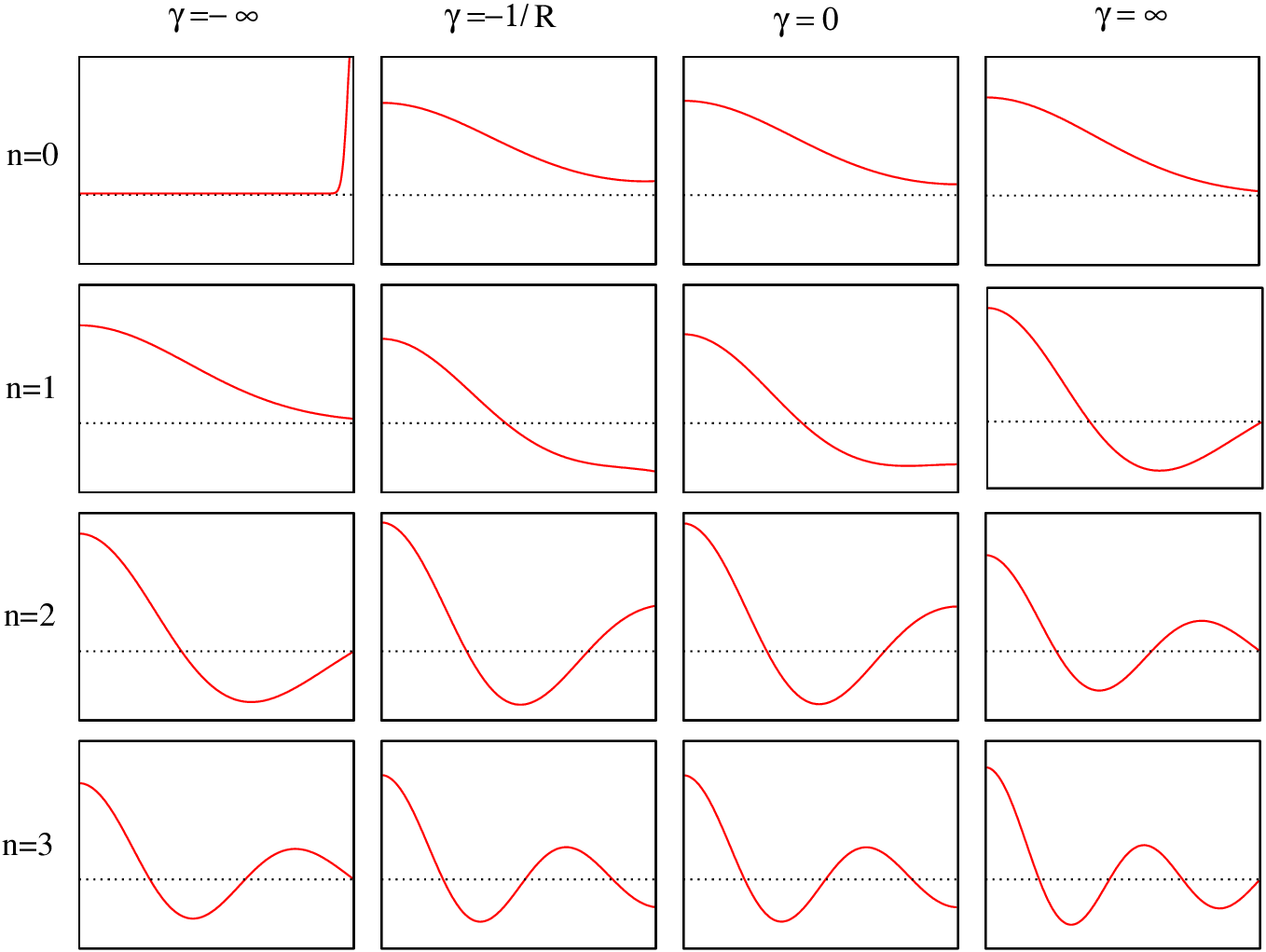,width=10cm}
\end{center}
\caption{\it Top: Spectrum of the of $m = 0$  states of the isotropic harmonic oscillator centered in a circular cavity
of radius $R =2.5 \alpha^{-1}$ with general boundary conditions as a
function of the self-adjoint extension parameter $\gamma$, rescaled to
$\arctan(\gamma R)$. The energy is given in units of $ \omega$.  The dotted
lines represent the spectrum for $\gamma = \infty$. Bottom: Wave functions of
the four lowest states with $n = 0, 1, 2,$ and $3$ for
$\gamma = \infty,0,-1/R$, and $-\infty$.}
\label{isofig5}
\end{figure}
\begin{figure}[tbh]
\begin{center}
\epsfig{file=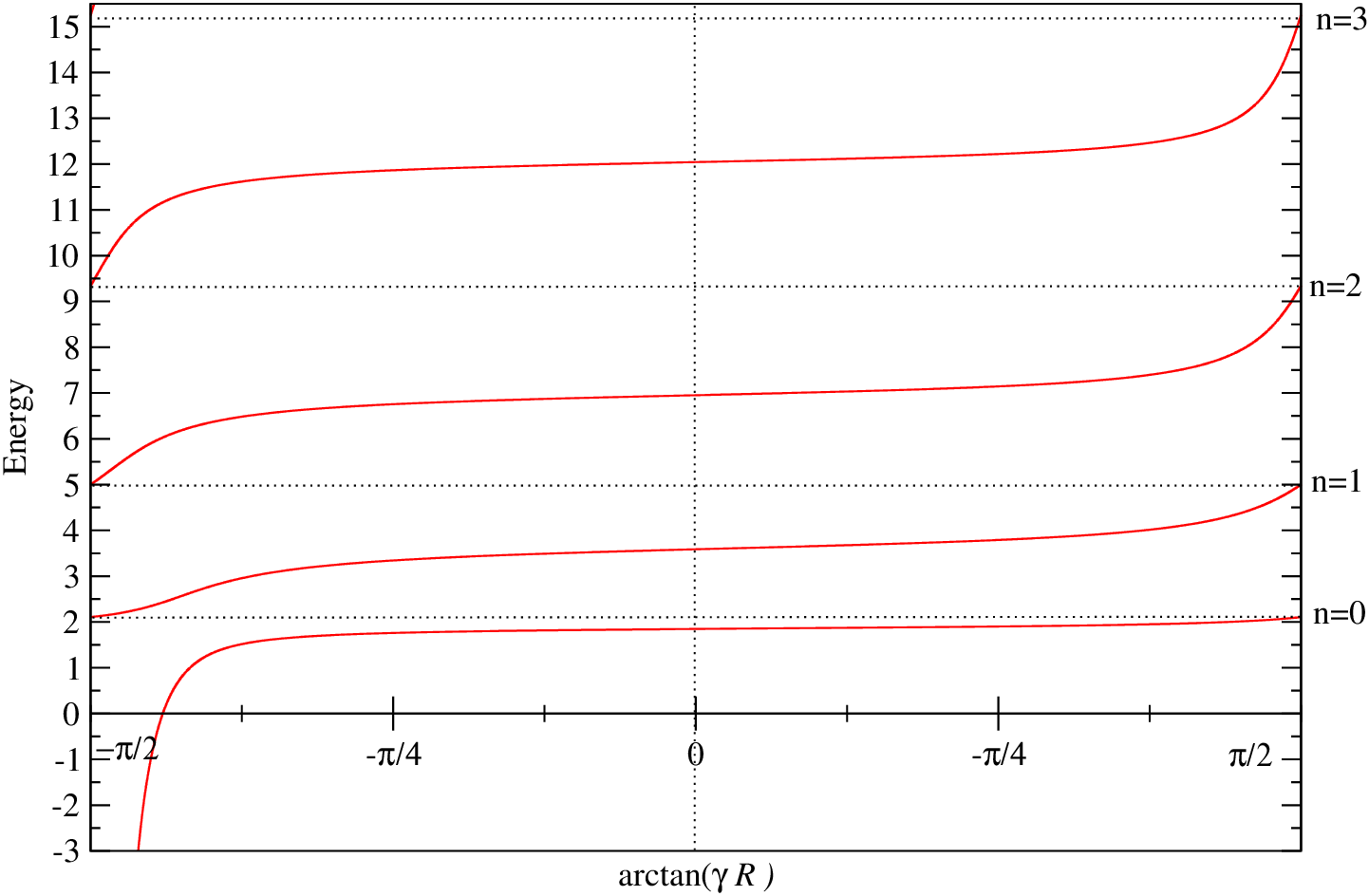,width=10cm} \vskip0.5cm
\epsfig{file=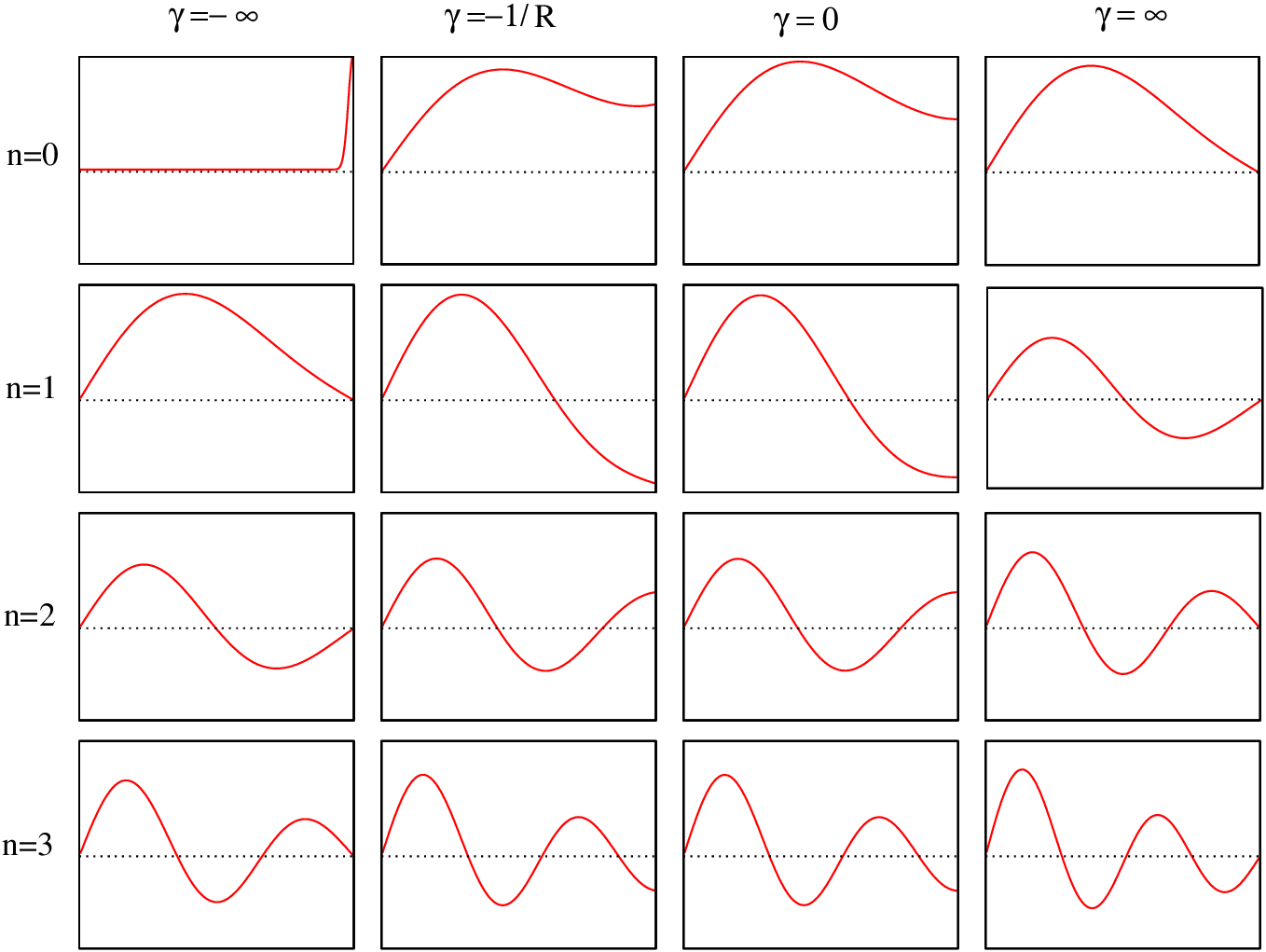,width=10cm}
\end{center}
\caption{\it  Top: Spectrum of the of $|m| = 1$  states of the isotropic harmonic oscillator centered in a circular cavity
of radius $R =2.5 \alpha^{-1}$ with general boundary conditions as a
function of the self-adjoint extension parameter $\gamma$, rescaled to
$\arctan(\gamma R)$. The energy is given in units of $\omega$.  The dotted
lines represent the spectrum for $\gamma = \infty$. Bottom: Wave functions of
the four lowest states with $n = 0, 1, 2$ and $3$ for
$\gamma = \infty,0,-1/R$, and $-\infty$.}
\label{isofig6}
\end{figure}
\begin{figure}[H]
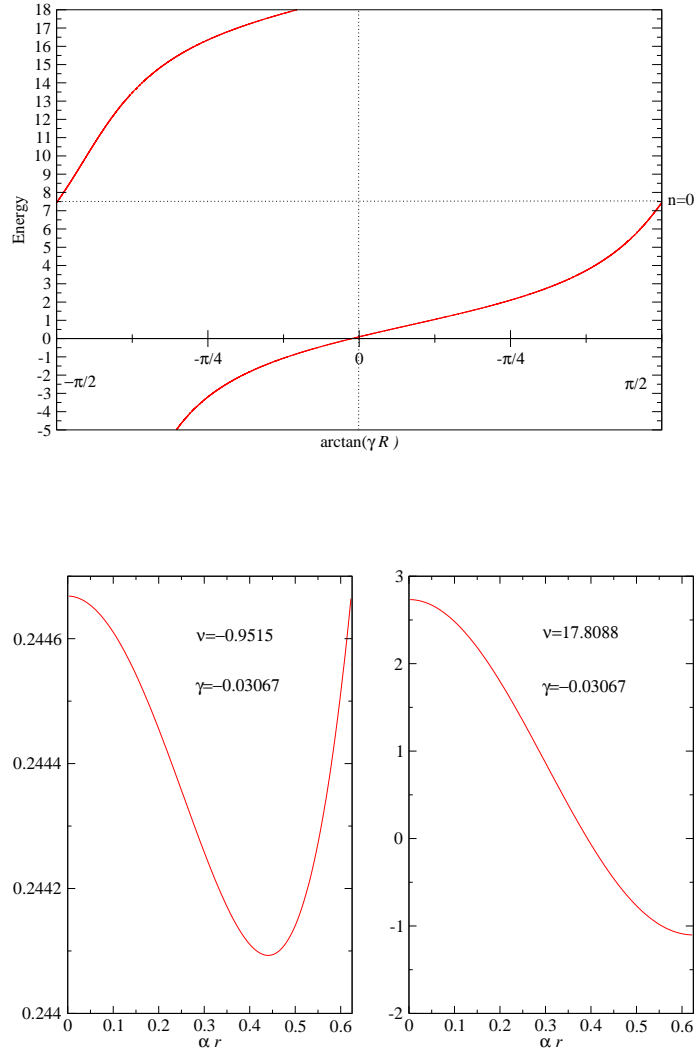

\begin{center}
\epsfig{file=isoC.eps,width=9cm} \vskip1.5cm
\epsfig{file=isoD.eps,width=9cm}
\end{center}
\caption{\it Top: Avoided level crossing between 0s and a 2s states for an isotropic harmonic oscillator centered in a circular cavity
of radius $R =0.625 \alpha^{-1}$ indicating a cavity resonance. The energy
is given in units of $\omega$. Bottom: Wave functions of the two states for
$\gamma = -0.0307 \alpha$, which are localized both near the center and at the wall
of the cavity.}
\label{isofig7}
\end{figure}

The finite volume effects on the energy spectrum are illustrated for the
standard Dirichlet boundary condition (with $\gamma = \infty$), and for Neumann boundary conditions (with $\gamma = 0$) in
figure \ref{isofig4}. In both cases, the accidental degeneracy between states of
different angular momentum, which is generated by the Runge-Lenz vector, is
removed. It is obvious that, in the case of Dirichlet boundary conditions, all the energies are shifted upward with the decreasing of $R$. On the other hand, the  Neumann
boundary conditions may lead to a downward shift of the energy for certain values of $R$, and to upward shifts for other values of $R$. The energy
spectrum for $m = 0$ as a function of the self-adjoint extension parameter
$\gamma$, and the corresponding wave functions of the states with
$n = 0, 1, 2, 3$ are illustrated in figure \ref{isofig5}.
Analogous results are shown in figure \ref{isofig6} for $|m| = 1$. The avoided level crossings
can be  noticed in these figures. It is more clearly visible in the higher energy levels. The reason for that is that the higher the energy,
the more likely is it that the particle interacts with the wall, the more likely is it that
bound states resonate with states localized at the cavity wall.  The fact that resonances in a finite volume manifest
themselves as avoided level crossings is familiar from quantum field theory, in
particular, lattice field theory \cite{Wie89,Lue91}. Figure \ref{isofig7}
(top) zooms in on an avoided level crossing between a 0s and a 2s state in a
circular cavity of radius $R = 1.25 \alpha^{-1}$. The corresponding wave functions are
illustrated at the bottom of figure \ref{isofig7}.
It is obvious from the bottom of figure \ref{isofig7} that a cavity resonance occurs for a  negative  value of $\nu$, which corresponds to a certain value of
$\gamma$. By using numerical methods, and calculating the values of $\gamma$ when the wave function shows a similar shape as the one in the bottom of figure \ref{isofig7}, for different cavity radii we obtain table 2.
\begin{table}[H]
\begin{center}
\begin{tabular}{|c|c|c|c|}

  \hline

 $R$& $\gamma$ & $\nu$ & $E$\\\noalign{\hrule height 2pt}
 4 & -3.4609 & -0.0099 & 0.9901 \\\hline
3.5 & -2.8662 & -0.0466 & 0.9534 \\\hline
3 & -2.2280 & -0.1450& 0.855\\\hline
2.5 & -1.5504 &-0.3196&0.6804\\\hline
2 & -0.8986& -0.5254&0.4746\\\hline
1.5 & -0.4070 & -0.7232&0.2768\\\hline
1 & -0.1240 & -0.8753&0.1247\\\hline
0.5 & -0.0157 & -0.9688&0.0312\\\hline

\end{tabular}
\end{center}
\caption{\it From the left: $R$ is the radius of the circular cavity in units of $\alpha^{-1}$, $\gamma$ is in units of $\alpha$, $\nu$ is the main quantum number, and $E$ to the right
is the resonance energy in units of $\omega$.}
\end{table}
\subsection{Integer $\nu$ for special values of $R$}
For the standard Dirichlet boundary condition (with $\gamma = \infty$), there are special values of $R$ that lead to  integers value of $\nu$. We will not cover this case here because it was studied in \cite{Sen1}. However, there are special values of $R$ that give integer values of $\nu$ in the case of general Robin boundary conditions. To find these values,  consider eq.(\ref{IsoSpectrum}). For $\nu=|m|$, the second term in eq.(\ref{IsoSpectrum}) vanishes. The confluent  hypergeometric function series in the first term terminates. Accordingly, we get the following relation
\begin{equation}
  |m|-R^2 \alpha^2+\gamma R=0
\end{equation}
The solution for the above equation gives
\begin{equation}\label{rInteger}
    R=\frac{\gamma +\sqrt{4|m|\alpha^2+\gamma^2}}{2 \alpha^2}.
\end{equation}
This means that if the radius of the cavity satisfies eq.(\ref{rInteger}) then there is one energy level with an  integer value $\nu=|m|$.\\
The other possibility is to take $\nu=2+|m|$. In this case, the confluent  hypergeometric function in the second term of eq.(\ref{IsoSpectrum}) is equal to $1$, while the confluent hypergeometric function series in the first term terminates. This leads to the following relation
\begin{equation}\label{rInteger2}
   -2R^2 \alpha^2+(1+|m|-R^2 \alpha^2)(|m|-R^2 \alpha^2+R \gamma)=0.
\end{equation}
The above equation is an algebraic equation of fourth order in $R$. The solution is available, but the expression is not very illuminating. However, there are cases for special values of $\gamma$ for which the expression for $R$ is representable. For example, when $\gamma=0$,  $R$ takes the following expressions
\begin{equation}\label{rInteger3}
    R=\frac{1}{\alpha\sqrt{2}}\sqrt{3+2|m|+\sqrt{9+8|m|}},\hspace{8mm}  R=\frac{1}{\alpha\sqrt{2}}\sqrt{3+2|m|-\sqrt{9+8|m|}}.
\end{equation}
Another special case that gives a short expression for $R$ is $\gamma=R\alpha^2$. For this case, $R$ takes the form
\begin{equation}\label{rInteger4}
    R=\frac{1}{\alpha}\sqrt{\frac{|m|(1+|m|)}{(2+|m|)}}.
\end{equation}
Accordingly, if the radius of the cavity obeys eq.(\ref{rInteger2}), or the special cases in eq.(\ref{rInteger3}), and  eq.(\ref{rInteger4}), then we have one energy level with integer main quantum number $\nu=|m|+2$.
\subsection{Self-Adjointness of the Runge-Lenz Vector}
For the isotropic harmonic oscillator in 2-d, there are three conserved operators.
The three operators are the one-component angular momentum operator, and the two components of the Runge-Lenz vector (technically speaking, it is a tensor). These operators generate
an $SU(2)$ symmetry. The components of the Runge-Lenz operator are
\begin{eqnarray}\label{RLiso}
R_x&=&\frac{1}{2M} \cos(2\varphi) \p_r^2
- \frac{1}{2 M r^2 } \cos(2\varphi) \p_\varphi^2 +
\frac{1}{M r^2 } \sin(2\varphi) \p_\varphi \nonumber \\
&-&\frac{1}{2} M \omega^2 r^2 \cos(2\varphi) -
\frac{1}{M r } \sin(2\varphi) \p_r \p_\varphi -
\frac{1}{2 M r} \cos(2\varphi) \p_r, \nonumber \\
R_y&=&\frac{1}{2M} \sin(2\varphi) \p_r^2
- \frac{1}{2 M r^2 } \sin(2\varphi) \p_\varphi^2 -
\frac{1}{M r^2 } \cos(2\varphi) \p_\varphi \nonumber \\
&-&\frac{1}{2} M \omega^2 r^2 \sin(2\varphi) +
\frac{1}{M r } \cos(2\varphi) \p_r \p_\varphi -
\frac{1}{2 M r} \sin(2\varphi) \p_r.
\end{eqnarray}
The accidental symmetry leads to additional degeneracies. Consider  a state in the infinite volume with quantum number $n$. Taking into account that energy levels have the same value when we replace $m$ with $-m$, an energy level with main quantum number $n$ has an $(n+2)$- fold degeneracy for even $n$, and an $(n+1)$- fold degeneracy for odd $n$. \\ In a finite volume,
when the isotropic harmonic oscillator potential is centered in the circular cavity, the domain of the Hamiltonian is subject to the condition in eq.(\ref{bcdot}). In order to investigate the symmetry of the system in this case, we must see which operators are still conserved. Let us first take  any component of the Runge-Lenz vector acting on the set of wave functions satisfying the condition in eq.(\ref{bcdot}) at $r=R$. When $R_x$ in eq.(\ref{RLiso}) acts on any wave function of eq.(\ref{ISOWF}) which is in $D(H)$, it transforms it into another function that does not satisfy eq.(\ref{bcdot}). This can be elucidated by the following.
\begin{equation}\label{Reffect}
    R_x \Psi_{\nu l}(r,\varphi)=\Phi_{\nu l}(r,\varphi),
\end{equation}
where $\Psi_{\nu l}(r,\varphi)$ satisfies the conditions in eq.(\ref{bcdot}). The expression for the new wave function $\Phi_{\nu l}(r,\varphi)$  is complicated. However, it can be proved that it does not satisfy the boundary condition in eq.(\ref{bcdot}). Accordingly, the operator $R_x$ maps the wave functions out of the domain of the Hamiltonian $H$. If we say that the components of the  Runge-Lenz operator are conserved because they commute with the Hamiltonian
\begin{equation}\label{RHcom}
    \frac{dR_x}{dt}=i[H,R_x]=i(HR_x-R_xH)=0,
\end{equation}
then this statement is meaningless in this case, because  $R_x$ maps the wave functions outside the domain of $H$, and thus $H$ cannot even act on the result. The same argument applies to the $R_y$ component.
This means that the symmetry associated  with the Runge-Lenz vector is broken in this case. \\ However, this is not the case for the angular momentum, because
\begin{equation}\label{LHcom}
    \frac{dL_k}{dt}=i[H,L_k]=i(HL_k-L_kH)=0,\hspace{3mm} k=1,2,3,
\end{equation}
 and this statement is still meaningful because
\begin{equation}\label{Leffect}
    L_z \Psi_{n l}(r,\varphi)=\Phi_{n l}(r,\varphi).
\end{equation}
Here $\Phi_{n l}(r,\varphi)$ still satisfies eq.(\ref{bcdot}), and therefore it is still in $D(H)$. Accordingly, the rotational  $SO(2)$ symmetry is still preserved.  As a result of this, for a certain energy level with main quantum number $n$, the degeneracy of the two states with $m$ and $-m$ is not removed. The previous argument implies that, in general, an energy level  $n$ splits into $(n+2)/2$ levels for even $n$, and $(n+1)/2$ for odd $n$ in a finite volume. This can be seen in figure \ref{isofig4}.\\ Here it is important to mention that in the case of the hydrogen atom in 3-d,  two fold application of the operator $R_+=R_x+iR_y$ on the wave function brings it back to $D(H)$, provided that $R=(l+1)(l+2)/2$, \cite{Sch00,Pup98,Pup02} which means that, there is an accidental symmetry under such special conditions. However, this is not possible in this case because of the expression of the Runge-Lenz vector given in eqs.(\ref{RLiso}), which contains a double derivative with respect to $r$. This makes it impossible to find a special value for $R$ that is energy-independent, under which $R_+^2$ returns the wave function into $D(H)$.
\section{Conclusions}
A free particle confined to a circular cavity with general perfectly reflecting boundary, characterized by a self-adjoint extension parameter $\gamma$, can have negative energy for negative values of $\gamma$. This is because it can be bound to the wall of the cavity. However, we proved that this does not violate the Heisenberg uncertainty relation when we used the version of this relation in a finite volume.
When the isotropic harmonic oscillator is placed in a circular cavity, the degeneracy of the energy levels associated with the Runge-Lenz vector is removed. On the other hand, the degeneracy associated with the rotational symmetry stays intact. The energy spectrum in our study takes a drastically different form than the one obtained  with the  standard Dirichlet boundary condition. It is especially characterized by an avoided level crossing when the particle is bound to the wall as well as to the center of force. This leads to a cavity resonance. The resonance energy has been calculated for different values of the cavity radius $R$. The same resonance arises in the case of the simple harmonic oscillator. \\The vibrational spectrum of confined molecules in a semiconductor structure can offer a possibility of verifying the influence of the self-adjoint extension parameter  on the vibration spectrum, for example, in an absorbtion frequency measurement, one can examine molecules confined to a cylindrical semiconductor structure with known radius and negligible hight relative to the radius.
\section*{Acknowledgments}
This work is supported in parts by the Schweizerischer Nationalfonds (SNF).
The author also likes to thank the city of Bern for their support in the
framework of the Swiss national qualification program
Bio\-medi\-zin-Na\-tur\-wis\-sen\-schaft-For\-schung (BNF). The ``Albert
Einstein Center for Fundamental Physics'' at Bern University is supported by
the ``Innovations- und Kooperationsprojekt C-13'' of the Schweizerische
Uni\-ver\-si\-t\"ats\-kon\-fe\-renz (SUK/CRUS).

\end{document}